\documentclass[twocolumn]{aastex631}

\begin{document}

\title{Why FLAMINGO is the perfect name for an array of Cherenkov telescopes}

\author{P. Flock} 
\affiliation{Flamingo International College, Indonesia}
\author{A. Laguna-Salina} 
\affiliation{Flamingo International College, Indonesia}
\author{F. James} 
\affiliation{Flamingo International College, Indonesia}
\author{G. Blossom} 
\affiliation{Flamingo International College, Indonesia}
\author{B. Carotene} 
\affiliation{Flamingo International College, Indonesia}
\author{C. Sparks} 
\affiliation{Flamingo International College, Indonesia}
\author{D. Tarek} 
\affiliation{Flamingo International College, Indonesia}
\author{A. Ahashia} 
\affiliation{Flamingo International College, Indonesia}
\author{J. Donald} 
\affiliation{Flamingo International College, Indonesia}
\collaboration{500}{(The \textcolor{magenta}{FLAMINGO} Collaboration)}
\email{spokesperson@flamingo.science}




\begin{abstract}
This paper argues why FLAMINGO (Fast Light Atmospheric Monitoring and Imaging Novel Gamma-ray Observatory) is the perfect name for an array of Cherenkov telescopes.
Studies which indicate pink is the most suitable pigment for the structures of Cherenkov telescopes have passed with flying colors. Pink optimizes the absorption and reflectivity properties of the telescopes with respect to the characteristic blue color of the Cherenkov radiation emitted by high-energy particles in the atmosphere. In addition to giving the sensitivity a big leg up, a pink color scheme also adds a unique and visually appealing aspect to the project's branding and outreach efforts. FLAMINGO has a fun and memorable quality that can help to increase public engagement and interest in astrophysics and also help to promote diversity in the field with its colorful nature. In an era of increasingly unpronounceable scientific acronyms, we are putting our foot down. FLAMINGO is particularly fitting, as flamingos have eyesight optimized to detect small particles, aligning with the primary purpose of Cherenkov telescopes to detect faint signals from air showers. We should not wait in the wings just wishing for new name to come along: in FLAMINGO we have an acronym that both accurately reflects the science behind Cherenkov telescopes and provides a visually striking identity for the project. While such a sea change will be no easy feet, we are glad to stick our necks out and try: FLAMINGO captures the essence of what an array of Cherenkov telescopes represents and can help to promote the science to a wider audience. We aim to create an experiment and brand that people from all walks of life will flock to.

\end{abstract}

\keywords{very-high-energy gamma rays; awesomeness; sparkling galaxies; non-thermal}


\section{Introduction} \label{sec:intro}
Cherenkov telescopes are a type of astronomical observatory that detect the Cherenkov radiation emitted by extensive air showers that are produced by high-energy cosmic rays or gamma rays interacting with the atmosphere. To achieve this, the telescopes use an array of mirrors that reflect the Cherenkov light onto a camera system that captures the images of these extensive air showers.

Cherenkov telescopes play a critical role in the study of high-energy cosmic rays and gamma rays, which often are produced by the most energetic phenomena in the universe. These particles carry important information about the sources and acceleration mechanisms of cosmic rays, as well as the properties of dark matter and other exotic astrophysical objects.

With the development of next-generation Cherenkov telescopes, scientists will be able to study these particles with greater sensitivity and accuracy than ever before. These instruments are expected to make major contributions to many areas of astrophysics, including the study of active galactic nuclei, gamma-ray bursts, and the search for dark matter. They will also enable scientists to study the universe at extremely high energies, shedding light on the properties of cosmic rays and the behavior of particles in extreme environments.

Finding a suitable name that encapsulates the essence of these missions is a crucial aspect of branding and public engagement efforts. In this paper, we present several reasons as to why FLAMINGO (Fast Light Atmospheric Monitoring and Imaging Novel Gamma-ray Observatory) is an excellent name for an array of Cherenkov telescopes.

Section~\ref{sec:res_disc} outlines and discusses the merit of FLAMINGO as a name for an array of Cherenkov telescopes, while Section~\ref{sec:summary} summarizes these ideas and arguments.

\section{Results \& Discussion} \label{sec:res_disc}
\subsection{FLAMINGO's color}
The color pink has been shown to be the most suitable color for Cherenkov telescopes. It optimizes the absorption and reflectivity properties of the telescope structure to enhance the detection efficiency of the predominately blue/UV Cherenkov light. Pink, being complementary to the blue Cherenkov light, provides the best properties to enhance visibility and improve the accuracy of the telescope's measurements.

The connection between the color pink and flamingos comes from the bird's distinctive coloration. Flamingos are known for their pink or reddish-pink feathers, which are caused by pigments in their food sources. These pigments, called carotenoids, are found in algae, crustaceans, and other aquatic organisms that the flamingos consume. Over time, the carotenoids accumulate in the flamingo's feathers, giving them their characteristic pink coloration \citep{flamingospink, yim2015}.
Given this association with the color pink, the name FLAMINGO provides a natural connection to this color.
\subsection{FLAMINGO's properties}
The vision system of flamingos is optimized for their feeding behavior adopted to filter-feeding, which allows them to detect and feed on very small particles \citep{flamingo_vision,flamingo_vision2}. Cherenkov telescopes are optimized for detecting faint flashes of light produced by high-energy cosmic rays and gamma rays as they interact with the Earth's atmosphere. These flashes are extremely brief and difficult to detect, requiring highly sensitive telescopes with advanced camera systems.
In addition, the visual sensitivity of birds, compared to humans, reaches lower wavelengths into the near-ultraviolet band \citep{flamingospink}, which is the dominant band to detect Cherenkov light.

Additionally, flamingos are known for their social behavior and their ability to communicate and coordinate with one another \citep{article}. Similarly, arrays of Cherenkov telescopes often operate with multiple telescopes working in tandem, allowing them to coordinate their observations and improve the accuracy of their measurements. Building and operating large arrays of telescopes requires are large amount of moeny and workpower. This can only be realized by forming collaborations of multiple research groups across the world. The name FLAMINGO, therefore, not only captures the scientific goals of an array of Cherenkov telescopes but also embodies the qualities of teamwork and coordination that are essential for its success.
\subsection{FLAMINGO's impact}
The name FLAMINGO has a fun and memorable quality that can help to increase public engagement and interest in the field of astrophysics. Using a name that is striking and easy to remember can help to capture the public's attention and generate excitement about the project.

\begin{figure}
\centering
\includegraphics[width=0.3\textwidth]{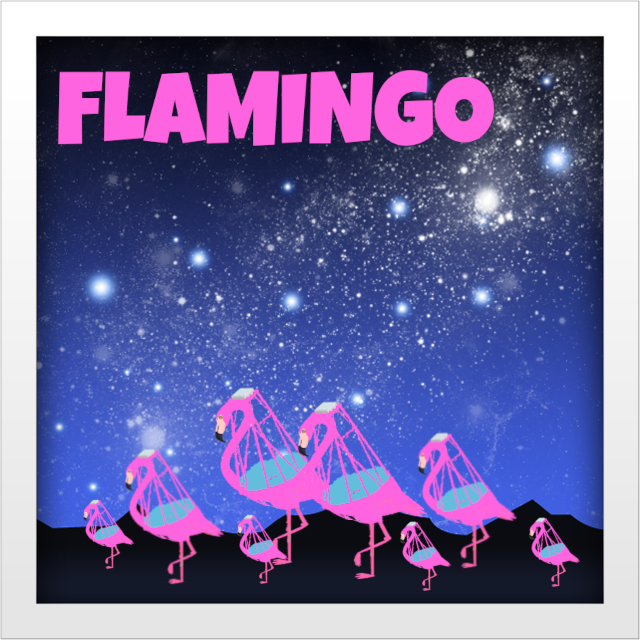}
 \caption{Preliminary logo design for the FLAMINGO collaboration.}
 \label{fig:prel_logo} 
\end{figure}

In addition, the name FLAMINGO can be used to create a unique visual identity for an array of Cherenkov telescopes. By incorporating images of flamingos into the project's branding and outreach materials, such as logos (see, e.g., Fig.~\ref{fig:prel_logo}), social media graphics, and promotional videos, the project can create a recognizable and distinctive brand that sets it apart from other astronomical observatories. 

The use of a memorable name and distinctive branding \citep{branding,boersma2018naming} can also help to increase public awareness and understanding of the science behind Cherenkov telescopes. By promoting the project through a variety of channels, including social media, press releases, and public outreach events, the project can engage with a broader audience and educate them about the fascinating science behind Cherenkov telescopes.

Finally, the name FLAMINGO can help to make arrays of Cherenkov telescopes more accessible and approachable to people who may not have a background in astrophysics. By using a name that is fun and easy to remember, the project can help to break down barriers. This, in turn, can help to generate interest in the field and inspire the next generation of scientists and researchers \citep{10.3389/fpsyg.2020.02088}.
\subsection{FLAMINGO's sensitivity}
\begin{figure*}
\centering
\includegraphics[width=12 cm]{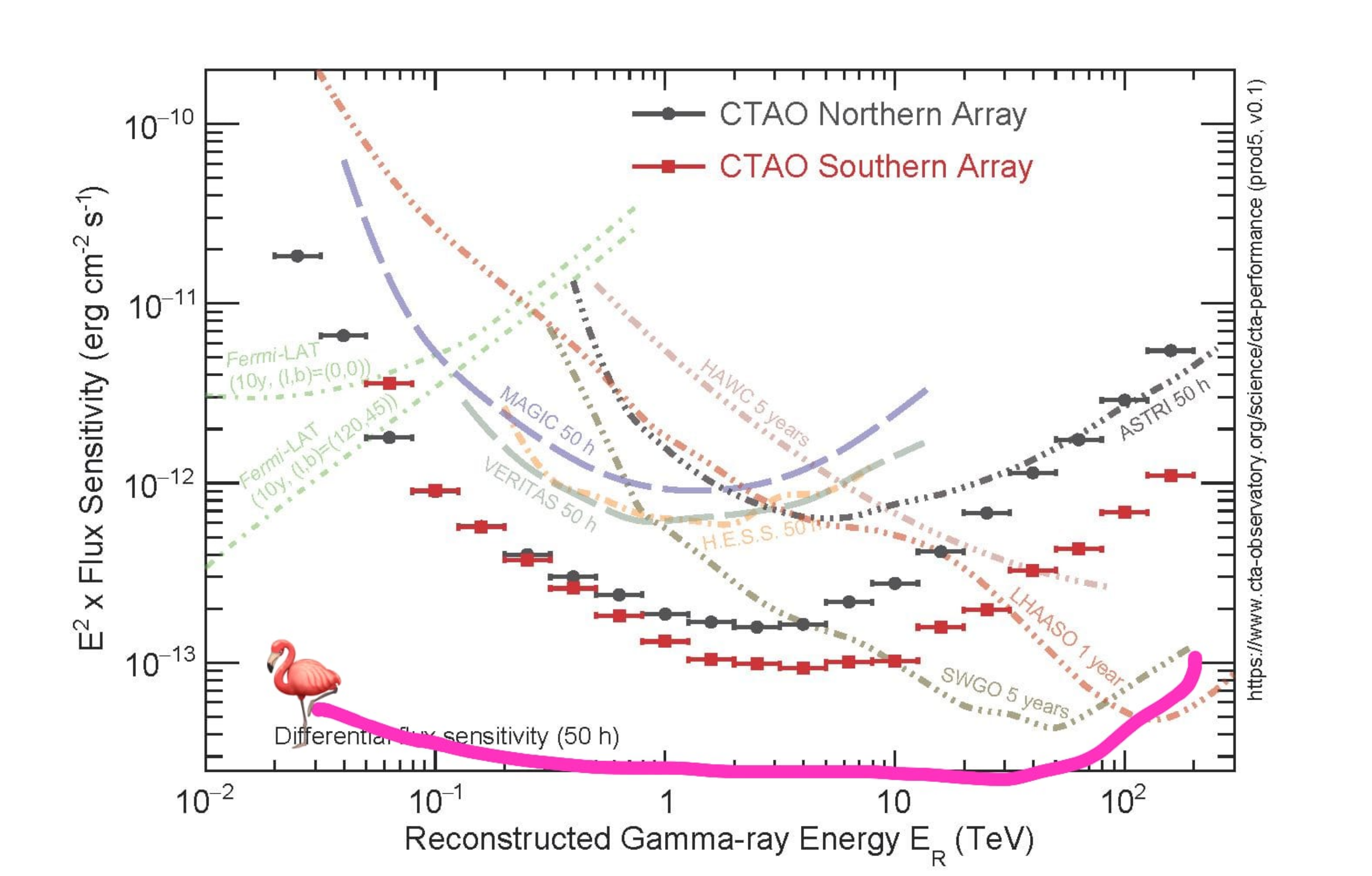}
 \caption{Differential sensitivity of FLAMINGO (solid pink line) with respect to other current or planned instruments. Figure adapted from \url{https://www.cta-observatory.org/science/ctao-performance/}.}
 \label{fig:sensitivity} 
\end{figure*} 
A structure painted in pink drastically enhances the sensitivity of imaging Cherenkov telescopes, allowing to reach an unprecedented performance in the very-high-energy gamma-ray range. In Fig.~\ref{fig:sensitivity} the differential sensitivity of FLAMINGO, obtained by highly sophisticated Monte Carlo simulations of the array configuration, is shown in solid pink in comparison to the sensitivities of other current and planned gamma-ray instruments. The improvement in sensitivity reached by a FLAMINGO array translates in the possibility to observe and study sources beyond the Cosmic Gamma Ray Horizon, up to a pinkshift of 10. Coloring not only the supporting structure of the telescopes of the array, but also the outside of the cameras in pink could further enhance the performance of the FLAMINGO array. Recent studies conducted by researchers of the FLAMINGO collaboration point to an even stronger performance if glitter is added to the pink painting. Glitter can also be applied moderately on the electronics and on the operators of the telescopes. 
This will be the object of a dedicated paper by the FLAMINGO collaboration, currently under preparation. We aim to further flatten the sensitivity curve of FLAMINGO, and reach a more fair and inclusive distribution of the sensitivity for all energies. On this regard, it is interesting to note that the current name of such curves could be misleading and should be actually changed to ``insensitivity" curves. As will be published in a dedicated study, an optimal sensitivity curve should be completely flat. 

\subsection{FLAMINGO's diversity}
Using the name FLAMINGO for an array of Cherenkov telescopes can also help to address diversity issues in the field of astrophysics. Astrophysics is one of the scientific fields that is still struggling with diversity and many demographic groups are underrepresented \citep{idea}. The use of an inclusive name and branding strategy can help to create a more welcoming and diverse community of astronomers and astrophysicists.

The use of a memorable name and distinctive branding can help to increase visibility for the project and attract a more diverse range of applicants for jobs and research positions. Moreover, by promoting the project through social media campaigns, public outreach events, and educational programs, the project can help to inspire and encourage a more diverse range of people to pursue careers in astrophysics.

In particular, for the LGBTQ+ community FLAMINGO has a high potential. The color pink carries a large weight for many members of the LGBTQ+ community, and today it is often used in LGBTQ+ pride flags and other symbols. By using a name and brand that recall the color pink, such as FLAMINGO, new experiments can help to promote awareness and support for the LGBTQ+ community. In addition, the use of colorful imagery can further emphasize the project's commitment to diversity and inclusion. A working group within the collaboration, \textit{FLAMINGO Pride}, is already tirelessly to help plan a number of events for Pride Month in June.

Overall, using the name FLAMINGO for an array of Cherenkov telescopes can help to create a more inclusive and diverse community of astronomers and astrophysicists, promoting a more equitable and representative field.

\section{Summary} \label{sec:summary}

In this paper, we have argued that FLAMINGO is the perfect name for an array of Cherenkov telescopes for several reasons. Firstly, the color pink, which is associated with flamingos, has been shown to be the most suitable color for Cherenkov telescopes, providing the best characteristics to enhance sensitivity. Secondly, the keen visual ability of flamingos aligns well with the primary purpose of Cherenkov telescopes to detect faint signals of high-energy cosmic rays and gamma rays. In addition, the name FLAMINGO has a fun and memorable quality that can help to increase public engagement and interest in the field of astrophysics. Using a name like FLAMINGO can also help to address diversity issues in the field of astrophysics by creating a more inclusive and welcoming community of astronomers and astrophysicists. 

In conclusion, the name FLAMINGO represents a powerful branding strategy for any next-generation array of Cherenkov Telescopes, capturing the essence of the scientific goals of the project while creating a fun and approachable image. By using the name FLAMINGO and associated branding, the project can generate excitement and engagement among the public while also inspiring and supporting a more diverse flock of astronomers and astrophysicists. Visit \url{https://flamingo.science}. to stay up to date on the ongoing efforts of the FLAMINGO collaboration.

\section{Authors' contributions}
Amazing Flamingo: paper editing, coordination of the project;
Brilliant Flamingo: investigation (Pink Matter searches with FLAMINGO which strongly supported the sensitivity study);
Colorful Flamingo: study of the sensitivity under different coloring of the structure;
Enthusiastic Flamingo:  endless positivity and enthusiasm and creation of an enjoyable atmosphere inside the collaboration;
Glittery Flamingo: investigation (glitter theoretical studies) and paper review;
Glowing Flamingo: study of the flattening of the sensitivity of the FLAMINGO array;
Harmonious Flamingo: Technical support and outreach activities, such as the composition of the FLAMINGO collaboration song;
Honorable Flamingo: guard of the FLAMINGO oath;
Incredible Flamingo: Software Board approval and test of the MonteCarlo simulations;
Radiant Flamingo: organization and supervision;
Sparkling Flamingo: review of the paper, simulations and visualization (Fig.~\ref{fig:sensitivity});
Visionary Flamingo: paper editing, FLAMINGO acronym development.

\vspace{0.5cm}
We would like to extend our sincere gratitude to everyone involved in the founding and development of the FLAMINGO project and all the currently active gamma-ray experiments. This includes the scientists, engineers, technicians, and administrative staff who have contributed their time, expertise, and resources to the project. Special gratitude goes to affordable beer and wine places, which laid the foundation for the creation of FLAMINGO.

Finally, we would like to acknowledge the many members of the FLAMINGO collaboration who have contributed to the development of the project. Thank you all for your hard work, dedication, and support of the FLAMINGO project. FLAMINGO is here to FLAMINSTAY.

%

\vspace{5mm}


\bibliography{sample631}{}
\bibliographystyle{aasjournal}



\end{document}